\documentclass{PoS}

\usepackage{amssymb,amsmath}
\usepackage{algorithmicx}
\usepackage{algpseudocode}
\usepackage{algorithm}
\usepackage{subcaption}

\usepackage{bm}

\providecommand{\innerp}[2]{#1^\dagger #2}
\providecommand{\norm}[1]{\lVert#1\rVert}
\newcommand{\bmin}{\begin{minipage}{0.32\textwidth}}
\newcommand{\emin}{\end{minipage}}
\newcommand{\bmini}[1]{\begin{minipage}{#1}}
\newcommand{\bc}{\begin{center}}
\newcommand{\ec}{\end{center}}

\title{ Split Grid and Block Lanczos Algorithm for Efficient Eigenpair Generation }
\ShortTitle{Split Grid and Block Lanczos}
\author{ Yong-Chull Jang and \speaker{Chulwoo Jung}  \\
Physics department, Brookhaven National Laboratory, Upton, NY 11973, U.S.A.\\
E-mail: \email{ypj@bnl.gov},
\email{chulwoo@bnl.gov} }
\abstract{
The increasing imbalance between the computing capabilities of individual nodes and the internode bandwidth makes it highly desirable for any Lattice QCD algorithm to minimize the amount of internode communication. One of the relatively new methods for this is the `Split Grid' or `Split Domain' method, where data is rearranged within the running of a single binary, so that the routines which requires significant off-node communications such as Dirac operators are run on multiple smaller partitions in parallel with a better surface to volume ratio, while other routines are run in one partition.

While it is relatively straightforward to utilize Split Grid method for inverters, the typical Lanczos algorithm which has one starting vector does not render itself naturally to Split Grid method. In this report we investigate the Block Lanczos algorithm(BL), which allows multiple starting vectors to be processed in parallel. 
It is shown that for a moderate number of starting vectors, BL 
achieves convergence comparable to similarly tuned Implicitly Restarted Lanczos algorithm (IRL)~\cite{Calvetti1994AnIR} on 2+1-flavor physical DWF/M\"obius ensemble.
}

\FullConference{The 36th Annual International Symposium on Lattice Field Theory - LATTICE2018\\
                22-28 July, 2018\\
                Michigan State University, East Lansing, Michigan, USA.}

\begin{document}

\section{Introduction}\label{sec:intro}
One of the characteristics of Lattice QCD(LQCD) which is different from other HPC applications is that the multi-node performance of LQCD applications is often limited by not only the network latency, but also the bandwidth, due to the nature of discretized grid Dirac operators which has relatively low computational intensity.
Even when ``Embarrassingly parallel'' approach is applicable, the need for memory  for intermediate data such as eigenvectors, propagators and  All-to-all(A2A) vectors often forces users to run on more than optimal number of nodes, or rely heavily on disk I/O.

Implicitly Restarted Lanczos~\cite{Calvetti1994AnIR} in combination with fine-tuned Chebyshev acceleration has been 
used by RBC/UKQCD collaborations successfully for some time.
The recent development of Multi-grid Lanczos~\cite{Clark:2017wom} has made it possible to generate even more eigenvectors without corresponding increase in memory requirement,
generating up to 8000 lowest-lying exact eigenvectors for 
2+1-flavor physical DWF/M\"obius emsembles
with the numerical cost similar to 20-30 undeflated inversions.

In addition to eliminating the critical slowing down as we approach phyical quark masses, exact eignvectors of preconditioned DWF/M\"obius Dirac operators has been used  to construct very accurate low-mode base for the All/Low Mode Averaging (AMA/LMA)~\cite{Blum:2012uh} and All-to-all~\cite{Foley:2005ac,Blum:2015you} methods.
However, the use of single starting vector makes IRL inherently serial, which limits the performance of the Lanczos kernel.

\begin{figure}
\bc
\includegraphics[width=0.8\textwidth]{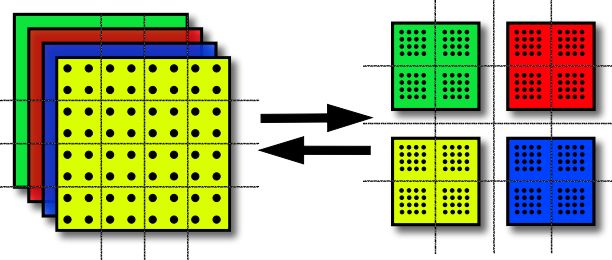}
\ec
\caption{Illustration of Split Grid algorithm \label{fig:SG}}
\end{figure}

Recently, RBC/UKQCD collaboration has started to explore the `Split Grid' method, where an application switches between running on one domain defined by single MPI communicator, and on multiple domains, often defined by split MPI communicators or something equivalent. Figure~\ref{fig:SG} illustrates the placement and movement of data in an application with Split Grid capability.
Split Grid  allows the communication intensive parts such as Dirac operator applications to run on multiple, smaller domains, while routines without much communications 
are done on the original domain. Split Grid acheived a significant speed-up in G-parity $K \rightarrow \pi\pi$ calculation reported in this conference~\cite{ckelly}, done on KNL based machines such as NERSC Cori in part.
The routines necessary for data rearrangement between the two domains have been implemented in Grid~\cite{Grid}, and as an standalone package using MPI RMA routines~\cite{BlockScramble}. For other fermion formulations where memory bandwidth is a limiting factor, block solver algorithms also can lead to a similar speed-up~\cite{deForcrand:2018orx}.

Here we explore utilizing the Split Grid method for Lanczos by switching to  Block Lanczos algorithm where there are multiple starting vectors.
A detailed description of the algorithm is presented in Section~\ref{sec:lanczos}. The performance and convergence of BL in comparison with IRL on 2+1-flavor physical ensemble is presented in Section~\ref{sec:24ID}. Summary and discussion is presented in Section~\ref{sec:conclusion}.

\section{Block Lanczos}\label{sec:lanczos}
Block Lanczos(BL) algorithm modifies the normal Lanczos by starting from 
$N_u(>1) $ orthonormal vectors 
$Y_0 = \{y_0^0, \cdots y_0^{N_u-1} \} $ and
calculate lowest $N_{stop}$ eigenvalues of  matrix 
\begin{equation}
A = T_n(X(M_{pc},\lambda_l,\lambda_h))\label{eq:op},
\end{equation}
where $M_{pc}$ denotes preconditioned Dirac matrix, $T_n(x)$ the Chebyshev polynomial of the first kind, and $X(x,\lambda_l,\lambda_h)$  is a linear function where $X(\lambda_l)=-1$ and $X(\lambda_h)=1$. 
Following a common notation for IRL~\cite{Calvetti1994AnIR}, $N_k(>N_{stop})$ is the number of initial Lanczos iterations before restarts, or convergence checks. $N_p$ denotes the number of the application of matrix between restarts, and $N_r$ is the number of restarts. It is assumed that both $N_k$ and $N_p$ are divisible by $N_u$.

Similar to the single-vector Lanczos, the basic equation for BL is
\begin{equation}
AY_{n-1} = Y_{n-2}\bm\beta^\dagger_{n-2} + Y_{n-1}\bm\alpha_{n-1} + Y_n \bm\beta_{n-1}
\end{equation}
The coefficient $\bm\alpha$ and $\bm\beta$ now are $N_u\times N_u$ complex matrix instead of real numbers and $\bm\beta$ is an upper triangular matrix
by construction via Gram-Schmidt processes. 

Besides the number of starting vector(s), the biggest difference between IRL and BL is the restarting strategy: in IRL, after each covergence check, a rotation matrix constructed from QR rotation matrices which suppresses eigenmodes with unwanted eigenvalues reduces the number of Lanczos vectors from $N_k+N_p$ back to $N_k$, so that it can repeat without continuing increase of memory usage. In BL studied here, all the Lanczos vectors are kept until the convergence is reached. While it can be in principle problematic, in practice it is not a concern: The distribution of eignevalues of a given Lattice QCD ensemble is extremely stable. This means a well tuned Chebyshev polynomial $T_n$ allows the convergence to be reached with a relatively small numeber of extra vectors, i.e. $N_p\times N_r \lesssim  N_k$.
Convergence is checked by calculating residual 
\begin{equation}
R_i = \norm{\left(M_{pc}- \langle v_i | M_{pc} | v_i \rangle \right)|v_i\rangle }.
\label{eq:res}
\end{equation} 

We also have studied a more direct extension of IRL in the name of 
Implicitly Restarted BL (IRBL)~\cite{Baglama2003IRBLAI}. We decided not to pursue this algorithm, as each rotation with QR destroys $N_u$ rows instead of 1, which means the degree of filtering polynomial available at each restart decreases by the factor of $N_u$, which slows the convergence significantly.

Despite the similarities between the 2 algorithms, there is a qualitative difference between the Krylov space generated:
\begin{equation}
{\cal K}_{n N_u}(A,y_0)=\mbox{span}\{y_0,A y_0,\cdots A^{n N_u-1} y_0\} \rightarrow  
\mbox{span} \{y_0,\cdots,y_{N_u-1}, A y_0,\cdots\,, A^{n-1} y_{N_u-1}\} \label{eq:space}
\end{equation}
As the Chebyshev acceleration is a critical part of the fast convergence of IRL for LQCD, 
Eq.~\ref{eq:space} suggests the number of matrix application for converging the same number of vectors will increase for a large enough $N_u$.
The details of BL is given in Algorithm 1.


\begin{algorithm}
\begin{algorithmic}[1]
\Procedure{$BL(A,\{\lambda_i,v_i\})$}{}
\State	$n \leftarrow 1, N_r \leftarrow 1$, $Y_0 = \{y_0^0, \cdots y_0^{N_u-1} \} $ 
\While {not converged}
\While {$ n < (N_k +N_r\times N_p) /N_u $}

        \State $\tilde{Y}_n = AY_{n-1}$ 
        \State For $n >1, \tilde{Y}_n \leftarrow \tilde{Y}_n - Y_{n-2}\bm\beta^\dagger_{n-2} $
        \State Orthogonalize $\tilde{Y}_n$  
against $\left\{  Y_1, \cdots Y_{n-2} \right\}$ (for numerical stability ) 
        \State $\bm\alpha_{n-1} = \innerp{Y_{n-1}}{\tilde{Y}_n}$ 
        \State $\tilde{Y}_n \leftarrow \tilde{Y}_n - Y_{n-1}\bm\alpha_{n-1}$ 
        \State Orthonormalize $\tilde{Y}_n$: $\tilde{Y}_n = Y_n \bm\beta_{n-1}$ 
        \State $n \leftarrow n+1$, 
\EndWhile
	\State Calculate eigenvector $v'_i$ of matrix 
\[
H_{n-1}=
\begin{bmatrix}
\bm\alpha_0 & \bm\beta^\dagger_0 & \cdots  & &  0 \\
\bm\beta_0  & \bm\alpha_1  & \bm\beta^\dagger_1 & \cdots \\ 
0           & \bm\beta_1  & \bm\alpha_2  & \bm\beta^\dagger_2 & \vdots  \\ 
\vdots      & \vdots      & \vdots       & \vdots   &    \vdots              \\
0           & \cdots      &\bm\beta_{n-3}  & \bm\alpha_{n-2}  & \bm\beta^\dagger_{n-2} \\
0           & \cdots      & 0                & \bm\beta_{n-2}  & \bm\alpha_{n-1}  
\end{bmatrix}
\label{eq:mat}
\]
\State Construct approximate eigenvectors $v_i$ of $A$ from $v'_i$ by 
$v_i = \bm{Y} v'_i, \bm{Y} = \left\{ Y_0,Y_1,\cdots Y_{n-1} \right\} $.
\State Check for convergence using Eq.~\ref{eq:res}.
 If less than $N_{stop}$ eigenvectors converged, $N_r \leftarrow N_r+1$.
\EndWhile
\EndProcedure
\end{algorithmic}
\caption{Block Lanczos \label{alg:BL}}
\end{algorithm}

\section{Performance of Block Lanczos on DWF $N_f=2+1$ ensemble on ALCF Theta }\label{sec:24ID}

Here we describe the performance and convergence of BL on 2+1-flavor physical DWF+ID ensembles with $V=(24^3, 48^3)\times 64\times 12, a\sim $0.2~fm~\cite{ID} which we refer as 24ID and 48ID ensembles respectively. 
While the evolution was done with M\"obius $Ls=24, b+c = 4$, RBC/UKQCD  has been using single precision zM\"obius with $Ls=12$ for the eigenvector generation.

For a realistic comparison, we start from the parameter RBC/UKQCD have been using for the generation of 1000 eigenvectors in fine grid, which forms the basis of Multi-Grid Lanczos~\cite{Clark:2017wom} for 2000 and 8000 eigenvectors for 24ID and 48ID ensembles respectively.
Since the convergence check on every Ritz vector is expensive and our experience with IRL suggest the convergence is almost always in increasing order in unaccelerated (without Chebyshev acceleration) eigenvalue or in decreasing order in accelerated Ritz value, we check the residual of every 32 Ritz vectors. Table~\ref{table:convergence} shows the number of converged eigenvectors after each convergence check.

\begin{table}
\begin{center}
\vspace{-2mm}
\begin{tabular}{c||c|c|c|c|c||c|c|c|c|c}
\hline
& \multicolumn{5}{|c||}{$24^3\times 64$ (24ID), 128 nodes } 
& \multicolumn{5}{|c}{$48^3\times 64$ (48ID), 512 nodes } \\
\hline
$N_u$ & 1(IRL) & 4 & 8 & 16 & 32 & 1(IRL) & 4 & 8 & 16 &32 \\
\hline
GF/node& 22	&53	&65	&88	& 99 & 36	&65	&76	& 85 &110 \\
\hline
$N_p\times N_r$&&		&		&	&		&	& 	 	& 	 	& 		& \\
\hline
320  &171	& 96 	& 0	 	& 0  & 0	 	&32	& 0	 	& 0	 	& 0		& 0 \\
480	 &621	& -		& -		& -	& -	&44	& 0	 	& 0	 	& 0		& 0 \\
640  &967	& 832 	& 704 & 576 & 128 	&61  & 32 	& 0	 	& 0		& 0 \\
800  &1040	& - 	&  -  & - 	& -  	&839 & 768 	& 704 	& 32	& 0 \\
960  &	&1056 & 1056 & 1024 & 864 &1040 & 928 	& 864 	& 736	& 608\\
1120 &	&	 	&     &   	&  -	&& 1120 	& 1088	& 928	& 736\\
1280 &	&		& 	& 		& 1056 	&& 	 	& 		& 1088 	& 964\\
1440 &	&		& 	& 		& 	 	&& 	 	& 		& 	 	& 1024\\
\hline
\end{tabular}
\end{center}
\caption{Performance of the Chebyshev-accelerated Dirac operator~(Eq.~\ref{eq:op}) in GFlops/node and the number of converged eigenvectors from single precision IRL and BL after each convergence check on DWF+ID ensembles. 
Numbers on the leftmost column below $N_p\times N_r$ are the total number of the operator application after initial Lanczos process for $N_k$ iteration.  
$N_k$ is 1040 for IRL and 1024 for BL. $N_p$ is 160 for both IRL and BL on 48ID, and 320 for BL on 24ID.
\label{table:convergence} }

\end{table}

\begin{figure}

\includegraphics[width=0.32\linewidth]{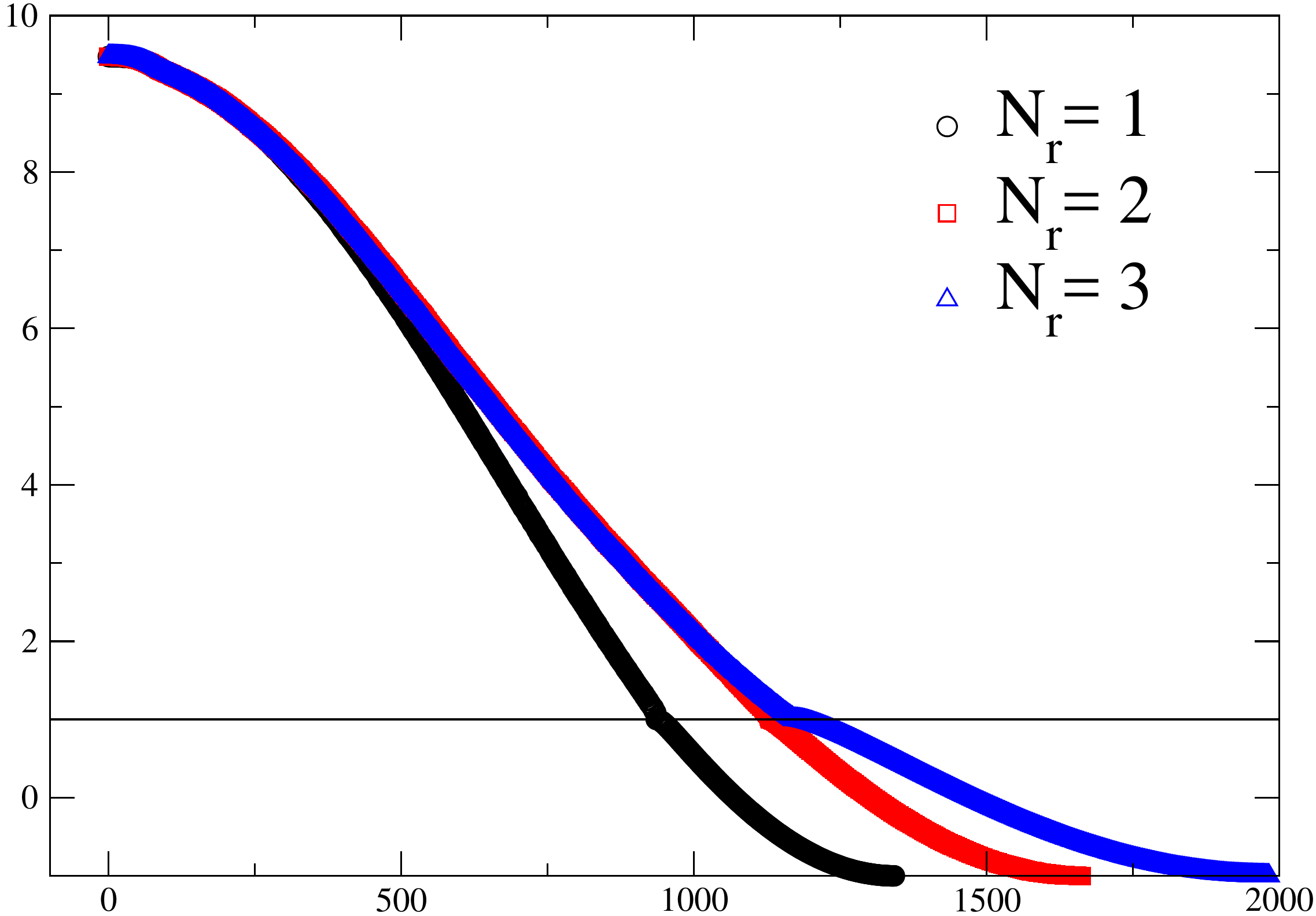}
\includegraphics[width=0.32\linewidth]{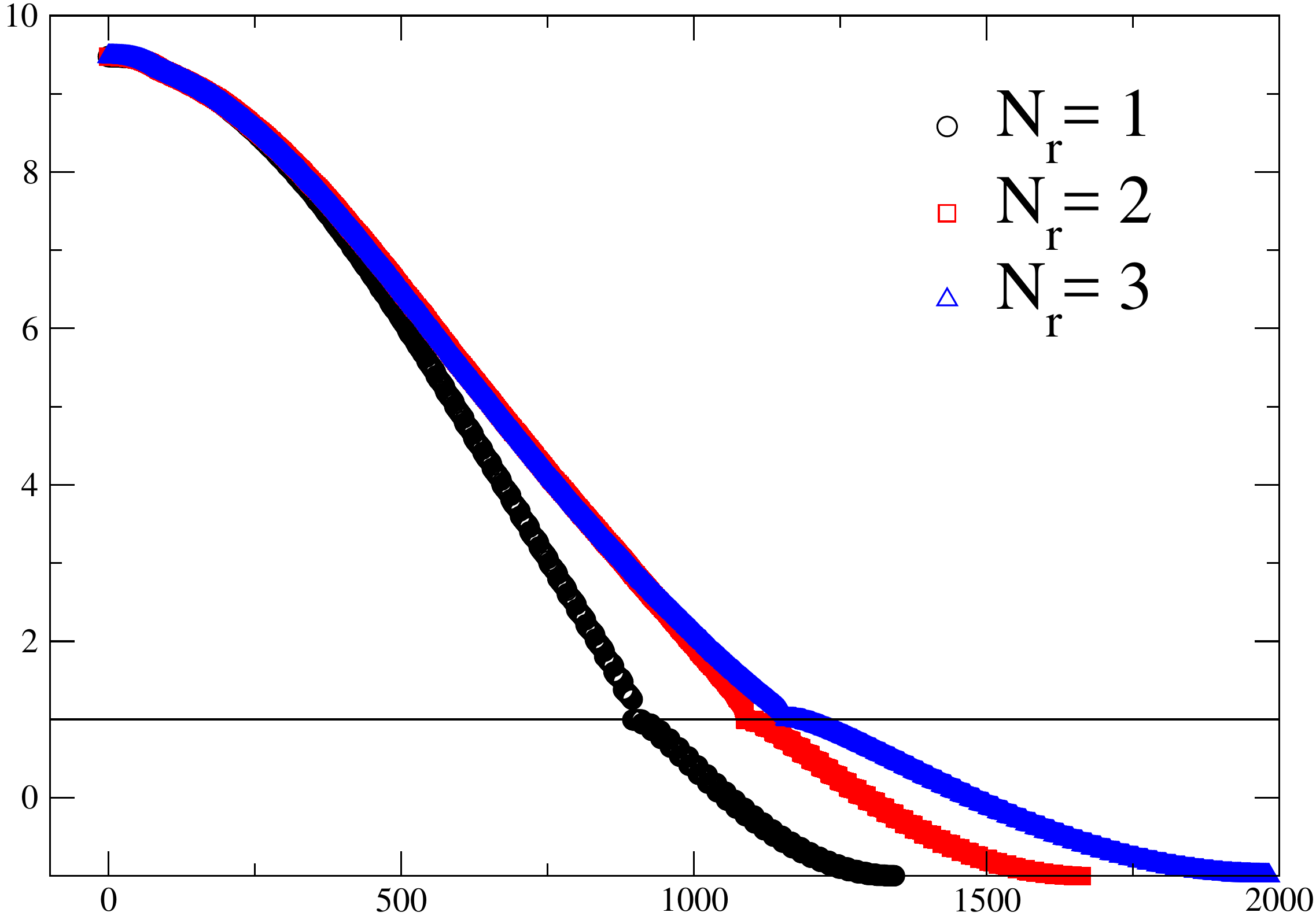}
\includegraphics[width=0.32\linewidth]{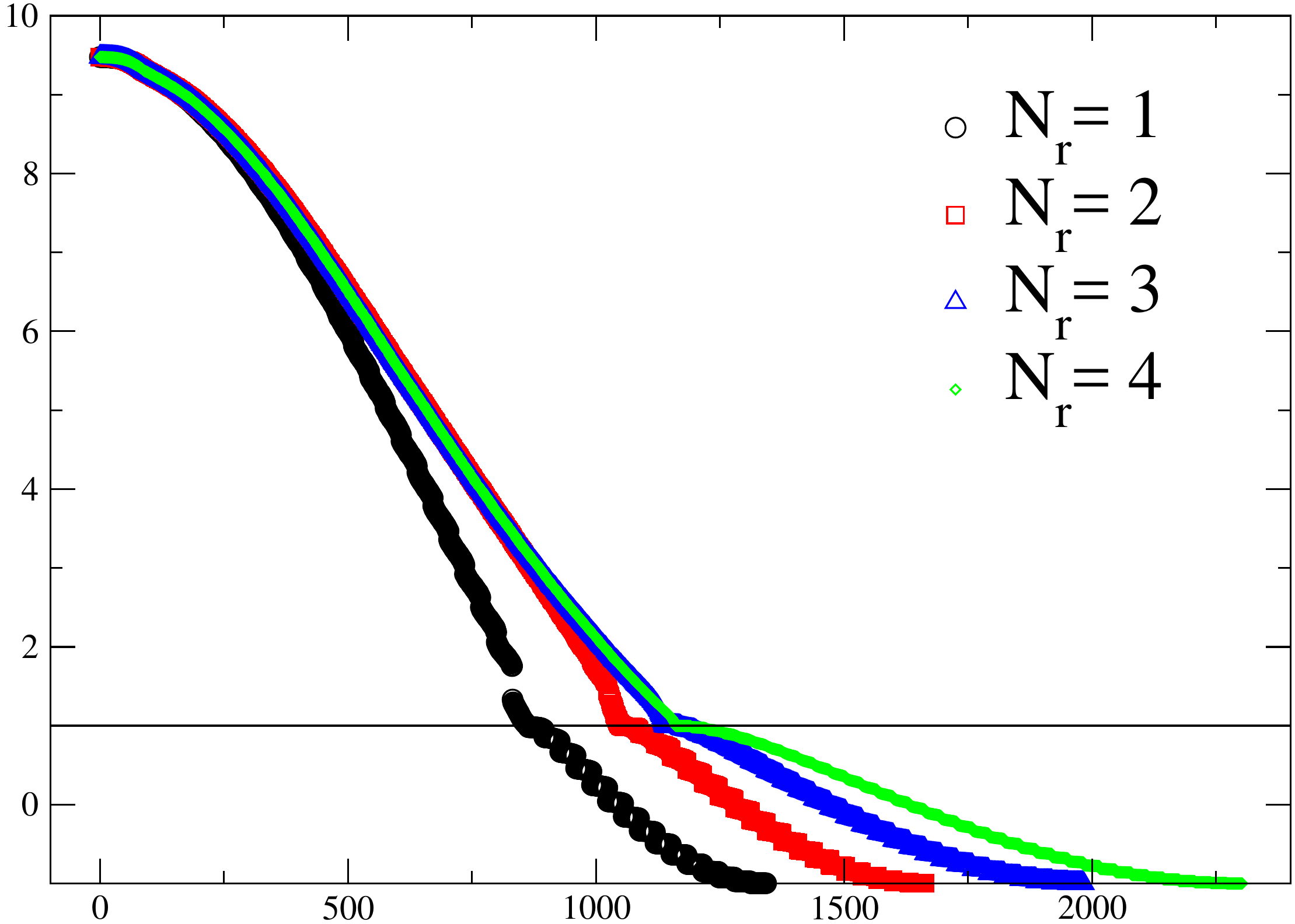}
\caption{Accelerated Ritz values $\langle v'_i | H_n | v'_i \rangle $ of Eq.~\ref{eq:mat} for a $24^3\times64, a\sim 0.2$fm 2+1f DWF configuration }
\end{figure}
\begin{figure}
\includegraphics[width=0.32\linewidth]{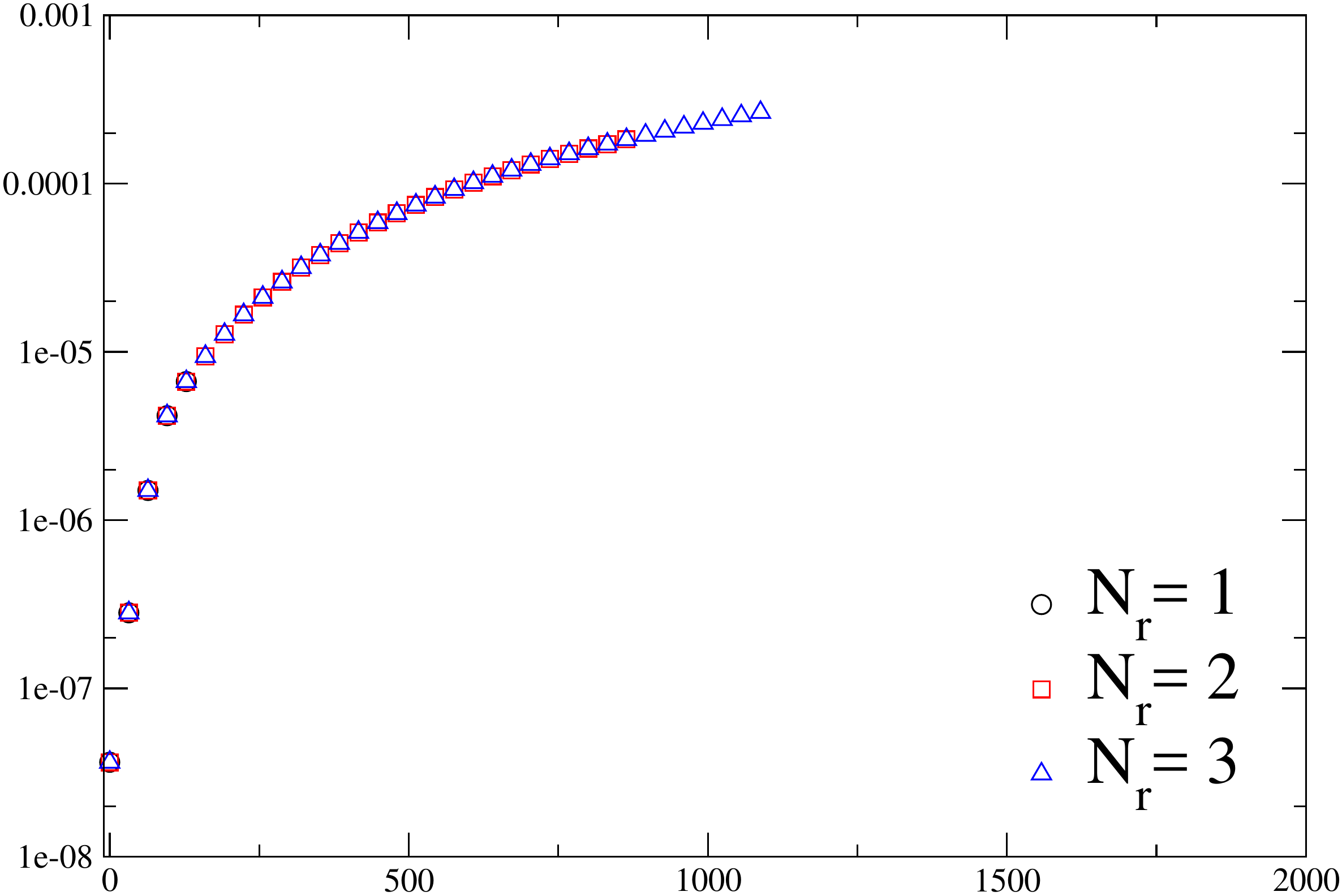}
\includegraphics[width=0.32\linewidth]{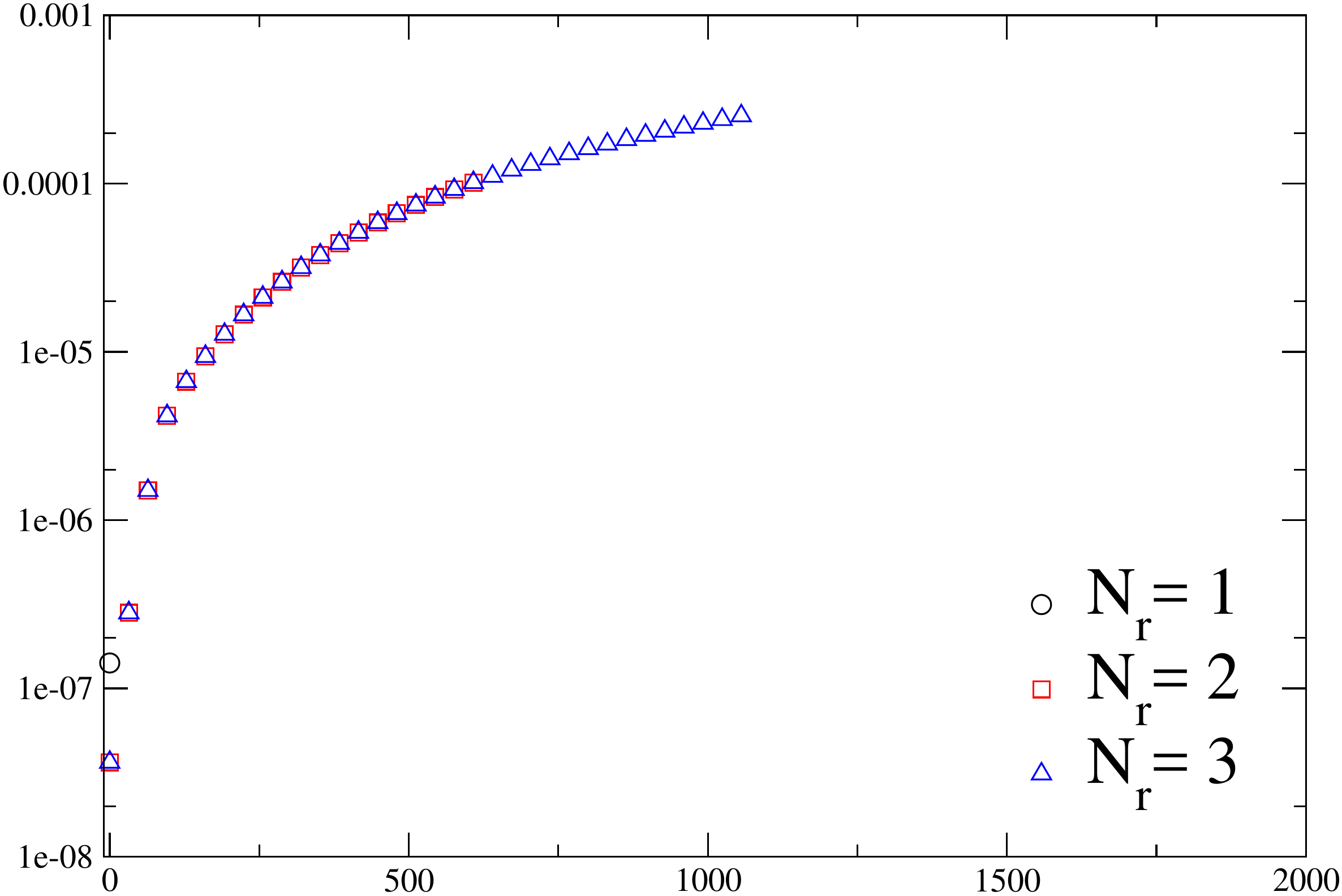}
\includegraphics[width=0.32\linewidth]{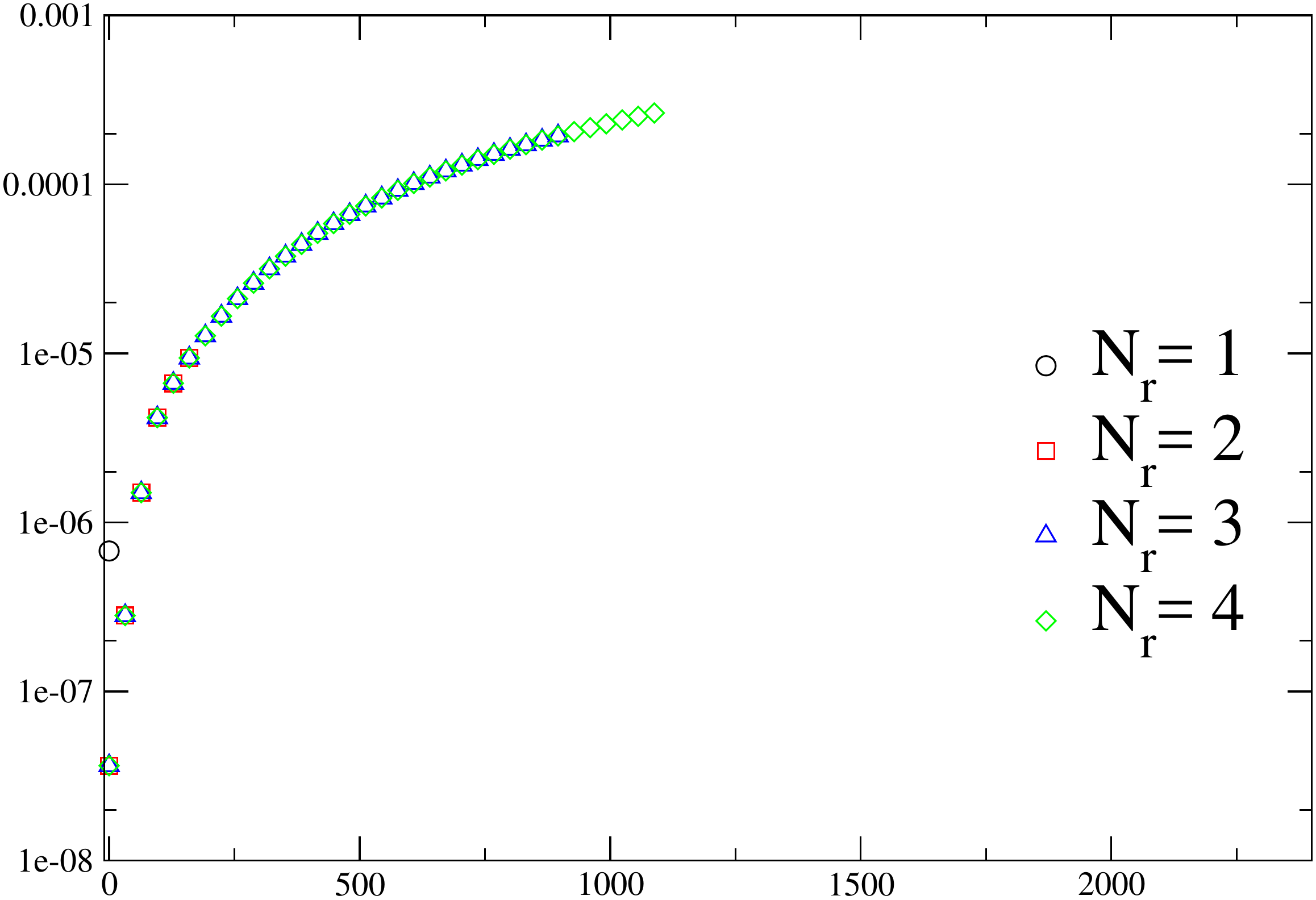}
\caption{
Unaccelerated Ritz values $\langle v_i | M_{pc} | v_i \rangle $
}
\end{figure}
\begin{figure}
\includegraphics[width=0.32\linewidth]{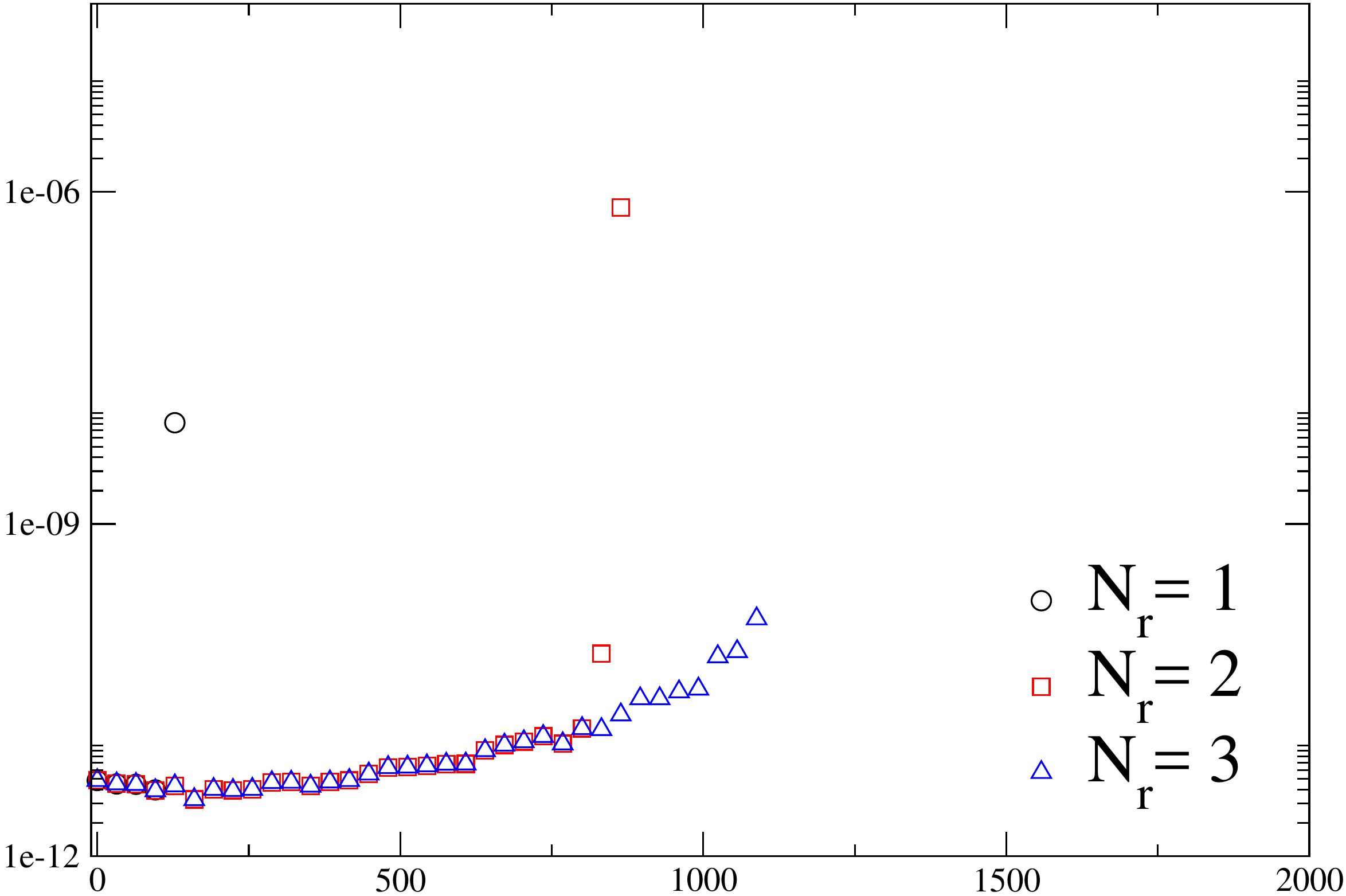}
\includegraphics[width=0.32\linewidth]{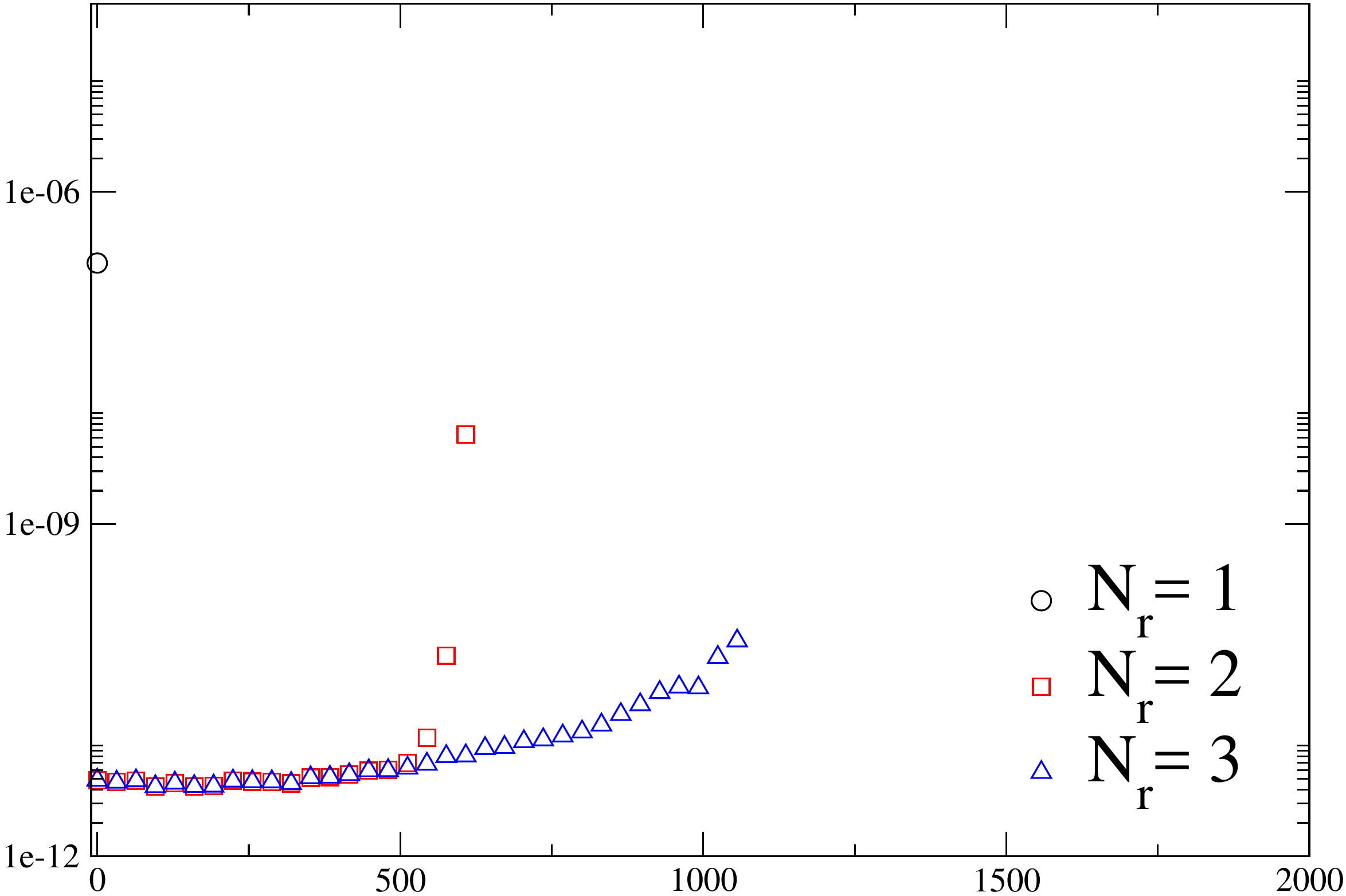}
\includegraphics[width=0.32\linewidth]{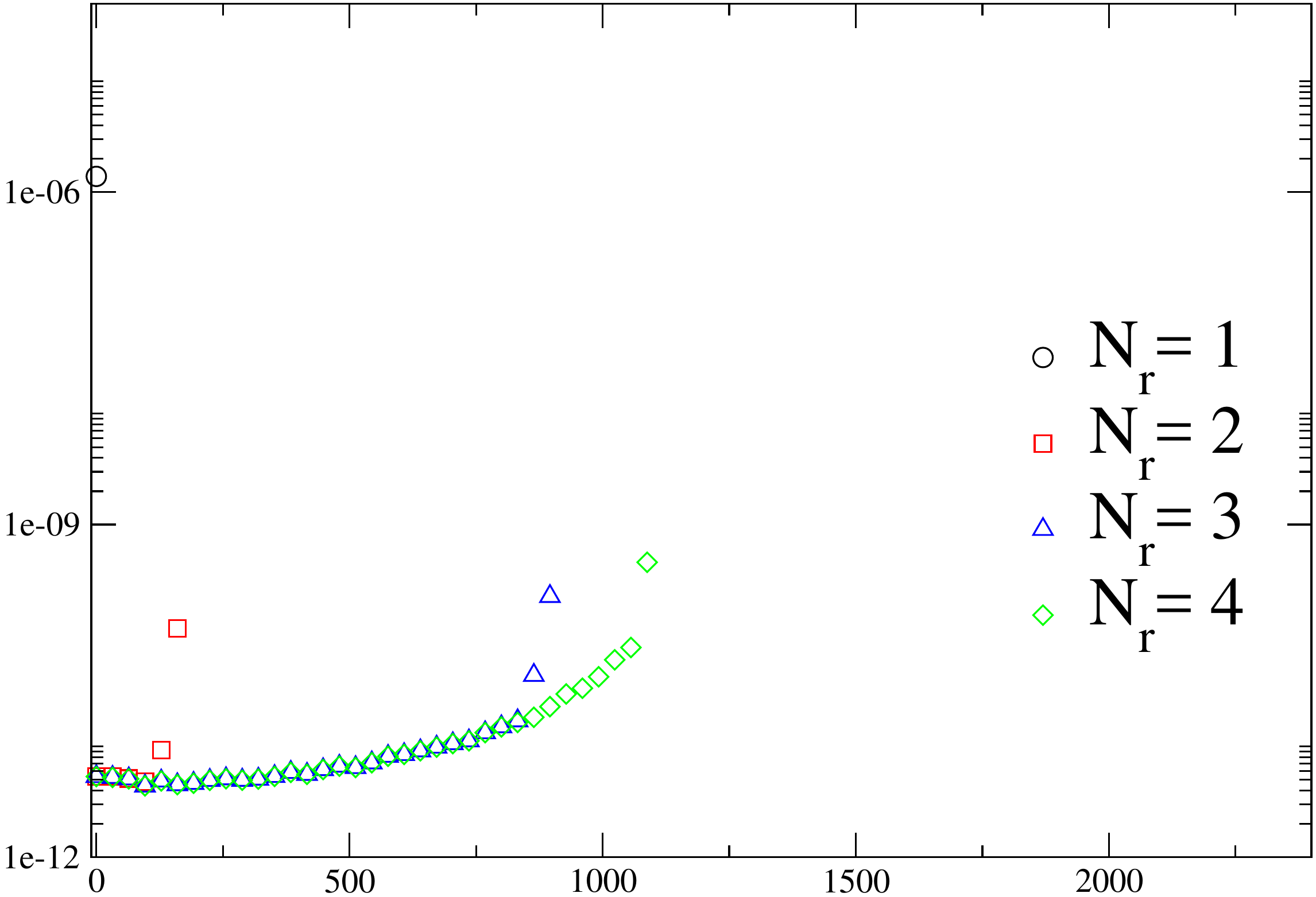}
\caption{ 
Residual of Ritz vectors (Eq.~\ref{eq:res}) at each convergence check for a Block Lanczos on a 24ID configuration~\cite{ID}.
}
\end{figure}
\begin{figure}
\bmini{.48\linewidth}
\bc
\includegraphics[width=\linewidth]{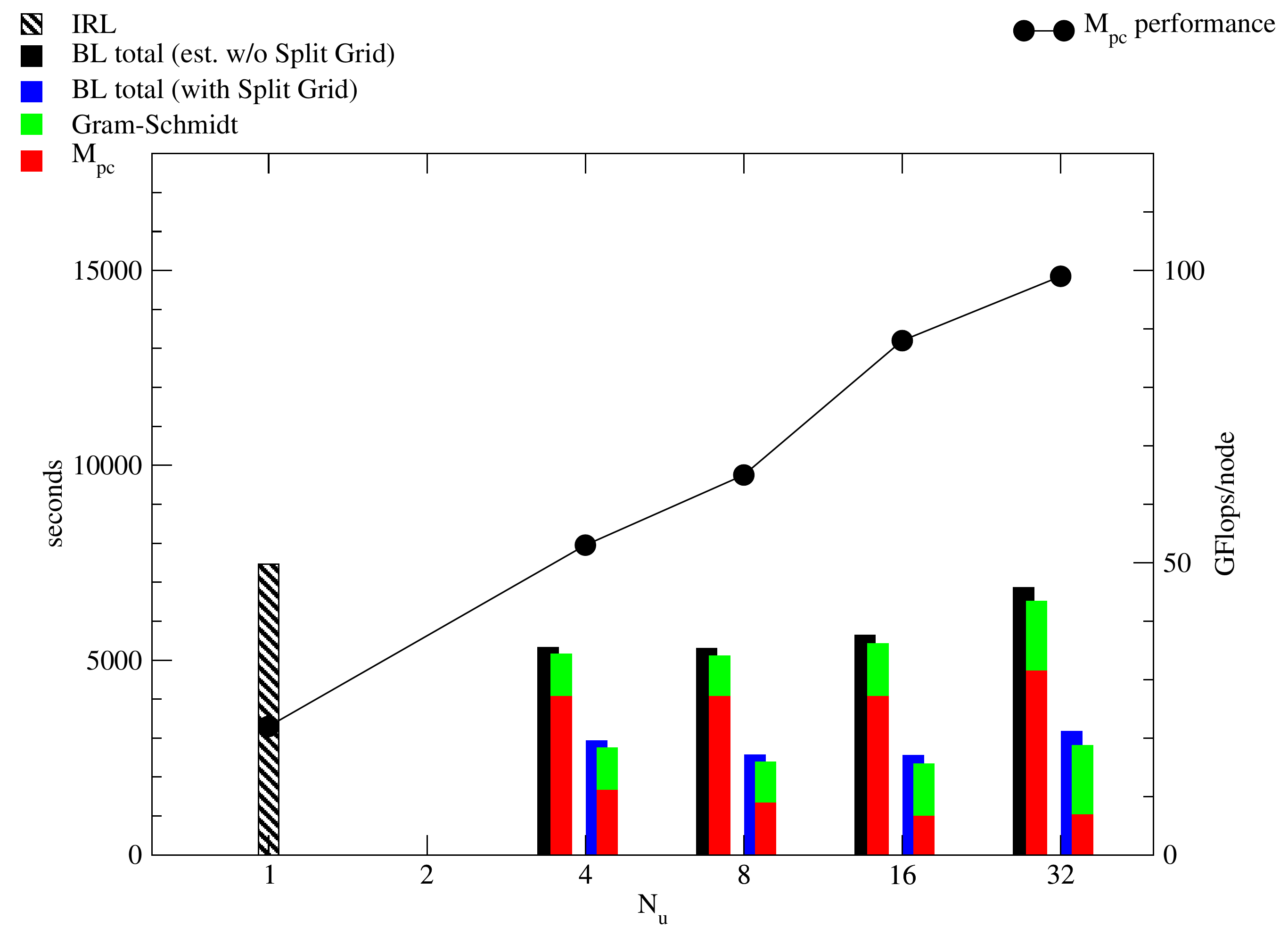}
24ID ensemble (128 nodes)
\ec
\emin
\vspace{0.04\linewidth}
\bmini{.48\linewidth}
\bc
\includegraphics[width=\linewidth]{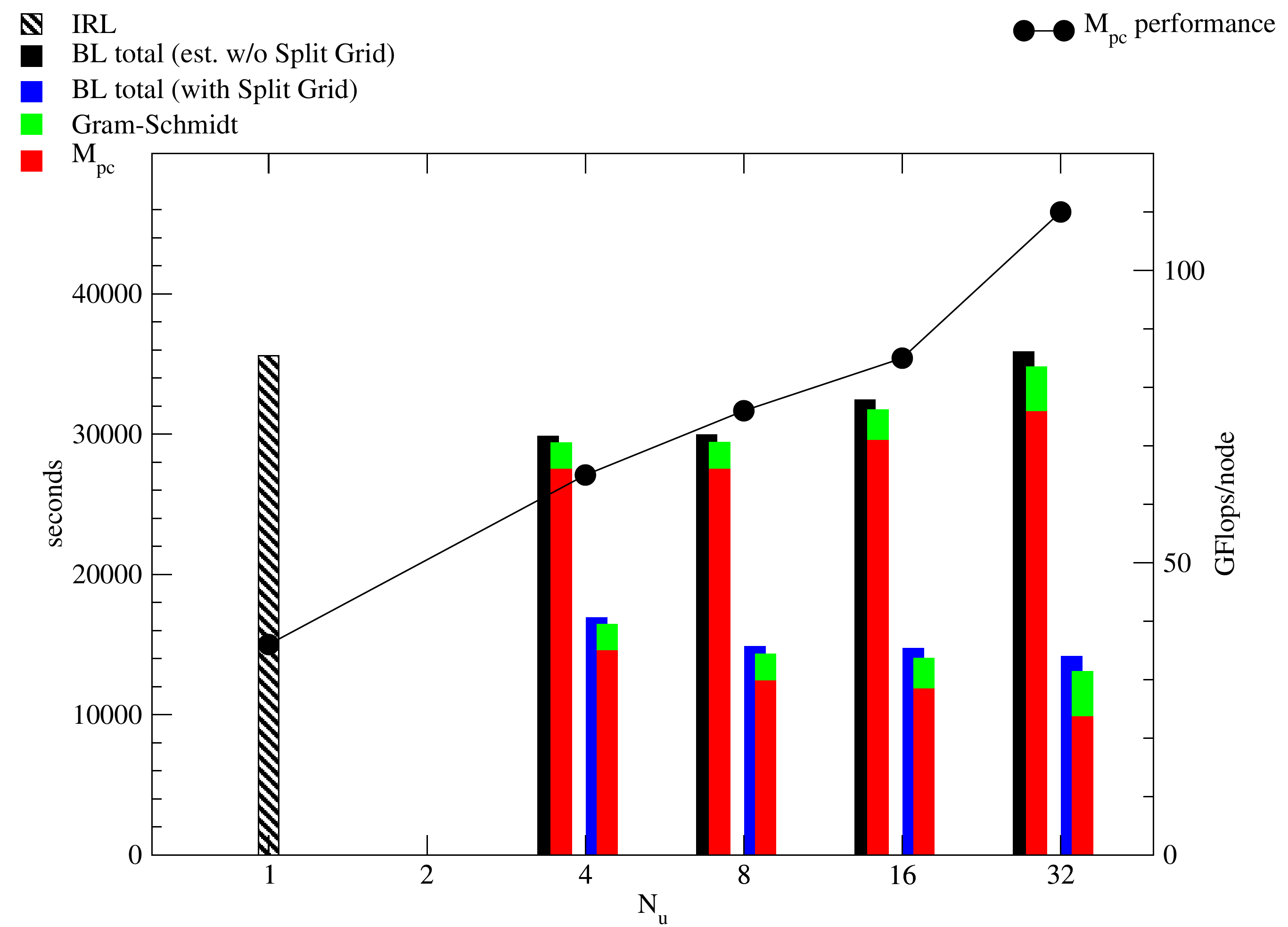}
48ID ensemble (512 nodes)
\ec
\emin
\caption{Performance of Chebyshev accelerated Dirac operaor $A=T_n(M_{pc})$ and timings of BL and IRL on 2+1-flavor DWF+ID ensemble, from Theta cluster(Cray XC40 with Intel Knights Landing CPUs)
at Argonne Leadership Computing Facility(ALCF).
\label{fig:timing}}
\end{figure}

Figures 2-4 shows the evolution of both accelerated and unaccelerated Ritz values and residuals after each restart. 
Figure~\ref{fig:timing} shows the breakdown of different part of BL compared to IRL run on the same number of nodes. 
The first bargraph for each $N_u$ is and estimate of timing for BL without split grid. They are estimates, as only the performance of Chebyshev kernel without Split Grid was measured, we can reasonably assume timings for the other parts of BL are the same  with or without Split Grid.

As also shown in Table~\ref{table:convergence}, the number of converged eigenvalues for a particular $N_r$ decreases as $N_u$ increases. 
However, the increase in time from the additional matrix application is more than well compensated by the performance increase from increasing local lattice volume. 
In fact, for the 48ID configuration, the total time is the smallest for the largest block size tried ($N_u=32$), and further increase in $N_u$ may still be beneficial.
It should be noted that the time spent on orthogonalization, shown as bars in green in Fig.~\ref{fig:timing}, 
 will grow larger relative to $M_{pc}$ for larger $N_{stop}$, dminishing gains relative to IRL. However, Gram-Schmidt and other linear algebra routines in BL can be further optimized by cache blocking or similar techniques made possible by  the presence of multiple Lanczos vectors. Further optimization is ongoing.

\section{Summary \& Discussion}\label{sec:conclusion}
Block Lanczos, in combination with Split Grid method, achieves a significant speed-up compared with the well tuned Implicitly Restarted Lanczos for the generation of lowest-lying eigenvectors of DWF/M\"obius 2+1-flavor physical ensemble on Intel Knights landing clusters. 
While the difference in the Krylov space generated by BL and IRL shown in Eq.~\ref{eq:space} 
negatively affects the convergence of BL, the increase in the number of Dirac operator application needed for convergence 
was moderate for up to $N_u=32$ for the DWF+ID physical ensembles. 
In fact, the most increase in the total runtime for $N_u>16$ was from the additional Gram-Schmidt necessary to keep the Lanczos vectors mutually orthogonal. 
Cache blocking and other well know techniques for local computation routines should extend the range of usability for BL.
It is also possible that a different tuning of acceleration polynomial could improve the convergence.

\section*{Acknowledgement}
This research was supported by the Exascale Computing Project (17-SC-20-SC), a collaborative effort of the U.S. Department of Energy Office of Science and the National Nuclear Security Administration.
This research used resources of the Argonne Leadership Computing Facility, which is a DOE Office of Science User Facility supported under Contract DE-AC02-06CH11357.
Block Lanczos investigated here is based on the Grid data parallel C++ mathematical object library~\cite{Grid}.



\begin{thebibliography}{1}

\bibitem{Calvetti1994AnIR}
D.~Calvetti, L.~Reichel and D.~C.~Sorensen, 
{\em Electronic Trans. Numer. Anal.} {\bf 2} (1994) 1.

\bibitem{Clark:2017wom}
M.~A. Clark, C.~Jung and C.~Lehner, 
{\em EPJ Web
  Conf.} {\bf 175} (2018) 14023 [\href{http://arXiv.org/abs/1710.06884}{{\tt
  1710.06884}}].

\bibitem{Blum:2012uh}
T.~Blum, T.~Izubuchi and E.~Shintani,
{\em Phys. Rev.} {\bf D88} (2013) 094503 
  [\href{http://arXiv.org/abs/hep-lat/1208.4349}{{\tt 1208.4349}}].

\bibitem{Foley:2005ac}
J.~Foley, K.~Jimmy~Juge, A.~O'Cais, M.~Peardon, S.~M. Ryan and J.-I. Skullerud,
{\em Comput. Phys.
  Commun.} {\bf 172} (2005) 145--162
  [\href{http://arXiv.org/abs/hep-lat/0505023}{{\tt hep-lat/0505023}}].

\bibitem{Blum:2015you}
T.~Blum, P.~A. Boyle, T.~Izubuchi, L.~Jin, A.~J{\"u}ttner, C.~Lehner,
  K.~Maltman, M.~Marinkovic, A.~Portelli and M.~Spraggs, 
{\em Phys. Rev. Lett.} {\bf 116} (2016), no.~23
  232002 [\href{http://arXiv.org/abs/1512.09054}{{\tt 1512.09054}}].

\bibitem{ckelly}
C.~Kelly, 
{\em PoS} {\bf LATTICE2018} (2018) 277.

\bibitem{Grid} https://github.com/paboyle/Grid

\bibitem{BlockScramble} https://github.com/chulwoo1/BlockScramble

\bibitem{deForcrand:2018orx}
P.~de~Forcrand and L.~Keegan, {\em Phys. Rev.} {\bf E98} (2018), no.~4 043306
  [\href{http://arXiv.org/abs/1808.01829}{{\tt 1808.01829}}].

\bibitem{Baglama2003IRBLAI}
J.~Baglama, D.~Calvetti and L.~Reichel, 
{\em SIAM J.
  Scientific Computing} {\bf 24} (2003) 1650--1677.

\bibitem{ID}
R.D.~Mawhinney and J.~Tu, 
{\em PoS} {\bf
  LATTICE2018} (2018).



\end{thebibliography}

\providecommand{\href}[2]{#2}\begingroup\raggedright\endgroup

\end{document}